\documentclass[epj]{webofc}
\usepackage[varg]{txfonts}   
\usepackage{braket}
\wocname{EPJ Web of Conferences}
\woctitle{CONF12}
\begin{document}
\selectlanguage{english}
\title{Deconfinement and chiral transition in AdS/QCD wall models supplemented with a magnetic field}
%
%

\author{David Dudal\inst{1,2}\fnsep\thanks{\email{david.dudal@kuleuven.be}} \and
        Diego R.~Granado\inst{3}\fnsep\thanks{\email{diegorochagranado@ufrrj.br}} \and
        Thomas G.~Mertens\inst{4,2}\fnsep\thanks{\email{tmertens@princeton.edu}}
}

\institute{KU Leuven Campus Kortrijk - KULAK, Department of Physics, Etienne Sabbelaan 51 bus 7800, 8500 Kortrijk, Belgium
\and
           Ghent University, Department of Physics and Astronomy, Krijgslaan 281-S9, 9000 Gent, Belgium
\and
           Departamento de F\'{i}sica, Universidade Federal Rural do Rio de Janeiro, BR 465-07, 23890-971, Serop\'{e}dica, RJ, Brasil
           \and Joseph Henry Laboratories, Princeton University, Princeton, NJ 08544, USA
}

\abstract{We discuss the phenomenon of (inverse) magnetic catalysis for both the deconfinement and chiral transition. We discriminate between the hard and soft wall model, which we suitably generalize to include a magnetic field. Our findings show a critical deconfinement temperature going down, in contrast with the chiral restoration temperature growing with increasing magnetic field. This is at odds with contemporary lattice data, so the quest for a holographic QCD model capable of capturing inverse magnetic catalysis in the chiral sector remains open.
}
\maketitle
\section{Introduction}
The ongoing heavy ion collision programs (ALICE or RHIC) provide a fertile testing ground for novel experimental, phenomenological and theoretical features of strongly coupled QCD. During the early stages after the collision, a sufficiently hot heat bath is created that leads to the creation of a quark-gluon plasma, an exotic state of deconfined QCD matter. In units of the fundamental QCD scale, $\Lambda_{QCD}$, the transition temperature $T_{deconf}\sim\Lambda_{QCD}$, so the relevant physics is still strongly coupled.

During the past decade, one witnessed an increased interest in the behaviour of (deconfined) QCD in the presence of a strong magnetic background field $\vec{B}$, where for reasons of simplicity one usually takes $\vec{B}=B\vec{e}_z$ with $B>0$ constant. The motivation for such studies is that during the early stages after a heavy collision, ultrastrong magnetic fields of the order of $eB \sim 1$-$15 m_\pi^2$ are expected to be created, with some evidence that $\vec{B}$ can be taken to be constant for the lifespan of the plasma phase \cite{Kharzeev:2007jp}. Let us refer to the recent review works \cite{Kharzeev:2013jha,Miransky:2015ava} for an overview of results and challenges. A particular interesting role for $B$ is reserved for unexpected (chiral) anomaly-induced transport phenomena, with possible analogues in condensed matter systems.
The presence of a preferred spatial direction in the plasma might also be relevant for explaining (observed) anisotropic behaviour in quark-gluon plasma related quantities. In particular, charmonia or transport properties in the plasma can acquire anisotropic behaviour, see for instance \cite{Dudal:2014jfa}.

In the current proceeding, we will mainly focus on the response of the 2 two standard QCD phase transitions to $B$: the deconfinement and chiral transition. For the chiral behaviour one intuitively expects the phenomenon of magnetic catalysis based on the dynamics of lower-dimensional systems \cite{Miransky:2015ava}. As the dynamics is expected to be ``squeezed'' along the $\vec{B}$-field for very strong magnetic fields, an effective reduction can be imagined, leading to similar results (including chiral magnetic catalysis) as in $1+1$-dimensional systems. Numerous papers appeared that indeed, at least initially, confirmed the magnetic catalysis, under the form of an increased chiral restoration temperature $T_c(B)$, see \cite{Kharzeev:2013jha,Miransky:2015ava} for references.

In 2011-2012, this commonly accepted lore was challenged by state of the art unquenched lattice simulations including a magnetic field which unequivocally showed the \emph{inverse} magnetic catalysis phenomenon and more precisely, a non-monotonous behaviour of the chiral condensate at finite temperature/magnetic field \cite{Bali:2011qj,Bali:2012zg}. A similar decreasing behaviour as for $T_c(B)$ was reported for the deconfinement temperature $T_{deconf}(B)$.  A few supporting theoretical works have appeared since then, see  \cite{Kharzeev:2013jha,Miransky:2015ava,Rougemont:2015oea} for useful references.

Next to lattice simulations, another powerful (analytical) tool to probe both static and dynamical questions in QCD is provided by the AdS/CFT \cite{maldacena} inspired description of strongly coupled gauge systems. The goal of this proceeding will therefore be to employ a modified AdS/QCD (soft) wall model to investigate whether the magnetic lattice QCD predictions for both deconfinement and chiral phase transitions can be corroborated from holographic AdS/QCD. We will report on our (ongoing) construction and tests of the magnetic AdS/QCD setup via lattice accessible quantities ($\sim$~non-dynamical questions), as once these issues are settled, our model can be exploited to study also quantities less amenable to lattice QCD ($\sim$~dynamical questions). For details we refer to our original work \cite{Dudal:2015wfn}.

\section{Our setup}
\subsection{General philosophy}
Holographic QCD is a specific application of gauge-gravity duality: a (classical) higher-dimensional AdS-gravity theory dual to a strongly interacting (quantum) gauge theory at the AdS  boundary in the $N_c=\infty$ limit. It can be seen as a generalization of the original correspondence formulated in \cite{maldacena}. Since strictly speaking it concerns an AdS/CFT duality, some leniency towards usage of ``CFT'' is required for people living in the (QCD) reality. Another working hypothesis is encoded in the observation that one expects that
\[\text{QCD}_{N_c=3}= \underbrace{\text{QCD}_{N_c=\infty}}_{\text{\tiny qualitative features}}+\underbrace{1/N_c~\text{corrections}}_{\text{\tiny quantitative corrections}}.\]
Since the advent of \cite{maldacena}, many holographic QCDish theories have seen the light: String-brane motivated models (e.g.~Sakai-Sugimoto \cite{Sakai:2004cn}), AdS wall models \cite{Erlich:2005qh,Karch:2006pv}, light front holography \cite{deTeramond:2005su}, \ldots.

In a sense, holographic QCD is a nice powerful ``geometrization'' of strongly coupled gauge dynamics. A downsize of the approach is that, usually, several constants need to be matched on top of external (QCD) input. In some cases, it is also not very clean as different choices of setup or parameter(s) can lead to quite different physics.

Confinement is modeled in/described by a suitable background metric (e.g.~AdS with a cut-off). This cut-off scale is needed to break conformal invariance, i.e.~to provide a QCD scale. Quark physics is mostly modeled in via probe branes/effective (Dirac-Born-Infeld) actions, corresponding to a holographic version of ``quenched QCD'' \cite{Karch:2002sh}. The latter is what makes it hard to capture all magnetic field effects: $B$ can only couple to the (neutral) glue/geometric background if the (charged) quark dynamics is taken into account. This is exactly the reason why it was (is) rather hard to study e.g.~the deconfinement transition in a $B$-field (amongst other things).
\subsection{The hard and soft wall AdS/QCD model}
The original holographic wall models \cite{Erlich:2005qh,Karch:2006pv} were effective descriptions for QCD dynamics, which necessitate the introduction of a mass scale playing the role of the IR QCD scale. In the AdS/CFT language the IR regime of the dual gauge theory penetrates the AdS bulk. Placing a wall in the AdS bulk serves to break the conformal invariance. This strategy allows to obtain a discrete mass spectrum. In this spirit, the hard wall model \cite{Erlich:2005qh} confines the AdS geometry to an infinite wall located at $r=r_c$ along the radial holographic coordinate. The soft wall model \cite{Karch:2006pv} rather introduces a smooth exponential fall-off along the holographic coordinate. This is achieved by assuming a dilaton $\Phi$ that vanishes at the UV boundary $r=0$ where the field theory (QCD) lives. In order to describe the QCD chiral dynamics both models work with an effective action whose fields are holographically dual to the left (right)-handed currents corresponding to the $SU(N_f)_R\times SU(N_f)_L$ chiral symmetry and to the chiral order parameter.

The AdS soft wall description of QCD as proposed in \cite{Karch:2006pv} relies on the action
\begin{equation}
\label{softwall}
S \propto\int d^{5}x \sqrt{-g} e^{-\Phi} \text{tr}\left[F_{L,\mu\nu}^2 + F_{R,\mu\nu}^2+|DX|^2-m_5^2|X|^2\right]
\end{equation}
where we refrained from writing the gravitational piece, viz.~the Einstein-Hilbert and Gibbons-Hawking boundary part. The corresponding metric reads
\begin{equation}
\label{ads}
ds^2 = \frac{L^2}{r^2}\left(-dt^2 + d\mathbf{x}^2 + dr^2\right)\,,\qquad e^{-\Phi} = e^{-c r^2},
\end{equation}
for the low $T$ confined phase. The coordinate $r$ runs from $r=0$ (the QCD boundary) to $r= \infty$, the AdS center. The high $T$ deconfined phase is encoded in the black hole metric
\begin{equation}
ds^2 = \frac{L^2}{r^2}\left(-f(r)dt^2 + d\mathbf{x}^2 + \frac{dr^2}{f(r)}\right) \,,\qquad
e^{-\Phi} = e^{-c r^2}\,,\qquad f(r) = 1 - r^4/r_h^4\,.
\end{equation}
Here we have $0\leq r\leq r_h$ with the Hawking temperature of the black hole determined by $T = \frac{1}{\pi r_h}$.

The hard wall predecessor of this soft wall model, \cite{Erlich:2005qh}, can be obtained by formally setting $c$ to zero while restricting $0\leq r \leq r_c$.
\subsubsection{Brief survey of the field content, chiral dynamics, deconfinement}
\begin{itemize}
\item The (left-handed) or (right-handed) flavour gauge field $A_{L,R}$ correspond to (gauge invariant) operators on the boundary (e.g.~left- and righthanded flavour currents whose 2-point correlaton functions encode vector or axial vector mesons).
\item $X$  is a scalar bifundamental field, with at leading order $X\sim\braket{\overline q q}$, the chiral condensate. $X$ also encodes the pion degrees of freedom.
\item One finds a string tension $\sigma_{QCD}\propto 1/r_0$ in the hard wall model. Unfortunately, there is no area law in the soft wall case \cite{Karch:2010eg}.
\item There is linear spectrum behaviour $m_n\sim n$ in the soft wall model, as expected from the Regge description of QCD. In the hard wall case, one rather has $m_n\sim n^2$.
\item The few model parameters are fixed by matching on QCD observables at $T=0$, with decent estimates for other states as an internal check.
\item There is QCD-desired chiral behaviour such as the Gell-Mann-Oakes-Renner relation.
\item The models \cite{Karch:2006pv,Erlich:2005qh} were proposed at $T=0$. In \cite{Herzog:2006ra} the soft and hard wall models were analyzed at finite temperature. Both models predict a deconfinement\footnote{The Hawking-Page phase transition is the holographic dual of the confinement/deconfinement phase transition. The idea is to compare the free energy in both thermal AdS geometry (confined phase) and AdS black hole geometry (deconfined phase) to obtain the critical temperature at which the transition occurs \cite{Witten:1998zw,Herzog:2006ra}.} phase transition at $T_{deconf}\sim 122$~MeV (hard wall) or $T_{deconf}\sim 191$~MeV (soft wall).
\item An important drawback of both models is that the metric does not solve the bulk Einstein EOMs. Though, more involved AdS/QCD models do, evidently at the cost of complication, already for $B=0$.
\end{itemize}


\subsubsection{A little more about the chiral properties}
According to the AdS/CFT correspondence \cite{Witten:1998qj}, one should have a boundary expansion of the form
\begin{equation}\label{rand}
X(r)= m r+ \sigma r^3+\ldots\,,
\end{equation}
where $m$ is the bare quark mass and $\sigma$ a measure for the chiral symmetry breaking (see later for a more precise statement).

In the hard wall model, $m$ and $\sigma$ are unrelated free parameters in the confining geometry. So there is no ``dynamical control'' over chiral symmetry breaking. In the deconfining geometry, one can show \cite{Dudal:2015wfn} that
\begin{equation}
\sigma\propto mT^2\,.
\end{equation}
So, there is actually never a chiral restoration ($\sigma\to0$) at any temperature $T$, except for the strict chiral limit.  For $m\neq0$, it is certainly not possible to identify deconfinement and chiral transition in the hard wall setting. The chiral phase transition in the soft model was first discussed in \cite{Colangelo:2011sr}.

Hence, there is a preference to use the soft wall model to probe the chiral transition, with or without $B$. We recall however that also the soft wall has an inherent ``chiral weakness'', namely that
\begin{equation}\sigma\propto m\,,\end{equation}
meaning there would be no more chiral symmetry breaking in the chiral limit, clearly a feature not shared by genuine QCD. It also implies that a relatively large bare quark mass $m$ is needed for realistic values of the chiral condensate, which eventually controls the mass of the pions in the $m\neq0$ case.

\subsection{Incorporating the magnetic field}
Having surveyed in brief the hard and soft wall models, we can now turn to a proper generalization thereof which includes a magnetic field. We base ourselves on the same philosophy of the original wall models: we depart from ``pure'' magnetized AdS (with or without black hole) and, as before, model a wall on top of it.  The magnetized AdS metric was constructed, analytically to leading order in $B^2$, in \cite{D'Hoker:2009mm}. The low $T$ metric reads
\begin{equation}
\label{thmetric}
ds^2=\frac{L^2}{r^2}\left(-f(r)dt^2+q(r)dz^2+h(r)\left(dx^2+dy^2\right)+\frac{dr^2}{f(r)}\right),
\end{equation}
with
\begin{eqnarray}
f(r)&=&1+\frac{2}{3}\frac{B^2r^4}{L^2}\ln \left(\frac{r}{\ell_c}\right)+\mathcal{O}(B^4)\,,\quad q(r)~=~1+\frac{8}{3}\frac{B^2}{L^2}\int_{+\infty}^{1/r}dx \frac{\ln(\ell_Y x)}{x^5}+\mathcal{O}(B^4)\,, \nonumber\\
h(r)&=&1-\frac{4}{3}\frac{B^2}{L^2}\int_{+\infty}^{1/r}dx \frac{\ln(\ell_Y x)}{x^5}+\mathcal{O}(B^4)\,,
\end{eqnarray}
while the high $T$ black hole metric now yields
\begin{equation}
\label{bhmetric}
ds^2=\frac{L^2}{r^2}\left(-f(r)dt^2+q(r)dz^2+h(r)\left(dx^2+dy^2\right)+\frac{dr^2}{f(r)}\right)\,
\end{equation}
with
\begin{eqnarray}
f(r)&=&1-\frac{r^4}{r_h^4}+\frac{2}{3}\frac{B^2r^4}{L^2}\ln \left(\frac{r}{\ell_d}\right)+\mathcal{O}(B^4)\,,\quad
q(r)~=~1+\frac{8}{3}\frac{B^2}{L^2}\int_{+\infty}^{1/r}dx \frac{\ln(r_h x)}{x^3\left(x^2-\frac{1}{r_h^4x^2}\right)}+\mathcal{O}(B^4)\,, \nonumber\\
h(r)&=&1-\frac{4}{3}\frac{B^2}{L^2}\int_{+\infty}^{1/r}dx \frac{\ln(r_h x)}{x^3\left(x^2-\frac{1}{r_h^4x^2}\right)}+\mathcal{O}(B^4)\,.
\end{eqnarray}
As before, the soft wall model is obtained by including into the action a dilaton factor $e^{-\Phi}=e^{-cr^2}$, while the hard wall is again given by cutting of AdS space at $r=r_c$. The metrics \eqref{thmetric} and \eqref{bhmetric} then corresponds to the confined, resp.~deconfined phase of the corresponding boundary QCD theory.

We draw special attention to the additional length scales $\ell_d$, $\ell_Y$ and $\ell_c$ that enter the metric form factors. Independent of these values, we have a solution of the Einstein EOMs. Though, $\ell_d$ and $\ell_Y$ will drop out of physical quantities, similarly to the AdS length $L$. Next to the Einstein EOMs, we must also require that there are no conical singularities in the Euclidean solution, which is the usual way to determine the Hawking temperature. These scales can then be exchanged for suitable functions of the other parameters. Importantly, in the confinement phase and thus also at the level of the deconfinement transition, $\ell_c$ does enter in physical quantities. As such, there is an extra relevant scale that needs to be fixed, for example by invoking $B\neq 0$ QCD (lattice) data.  In the earlier work \cite{Mamo:2015dea}, $\ell_c=1$ was tacitly assumed from the beginning.

Let us enlist a few more technical results, for ample details we refer to \cite{Dudal:2015wfn}:
\begin{itemize}
\item For the magnetic field, we have $B\equiv B^{5D}\neq B^{4D}\equiv\mathcal{B}$, but a sensible identification based on bulk-boundary matching learns that $\mathcal{B}=1.6\frac{B}{L}$.
\item The Ricci curvature is given by $R=-\frac{20}{L^2}+\frac{2}{3}\frac{r^4}{L^4}B^2$, both for \eqref{thmetric} and \eqref{bhmetric}. So, there is a singularity in the deep interior for $B\neq0$. Luckily, including a wall (by hand) ``hides'' this naked singularity.\\The above metric is valid if $B^2<T^2$ (confined)  and if $B^2r^4<L^2$ (deconfined). As such, the large $r$ region is actually not accessible perturbatively.  We will anyhow assume that $\mathcal{B}$ (and thus $B$) is sufficiently small, as this is the area of phenomenological interest.
\item The black hole metric has a rather involved (double, sometimes degenerate) horizon structure if $B\neq 0$. To make a long story short, given a $T$ and $\mathcal{B}$, all parameters of the (black hole) metric can be determined unambiguously, up to $\ell_c$.
\end{itemize}

\section{Our results}
\subsection{The deconfinement transition}
\begin{figure}[h]
  \centering
   \includegraphics[width=0.5\textwidth,clip]{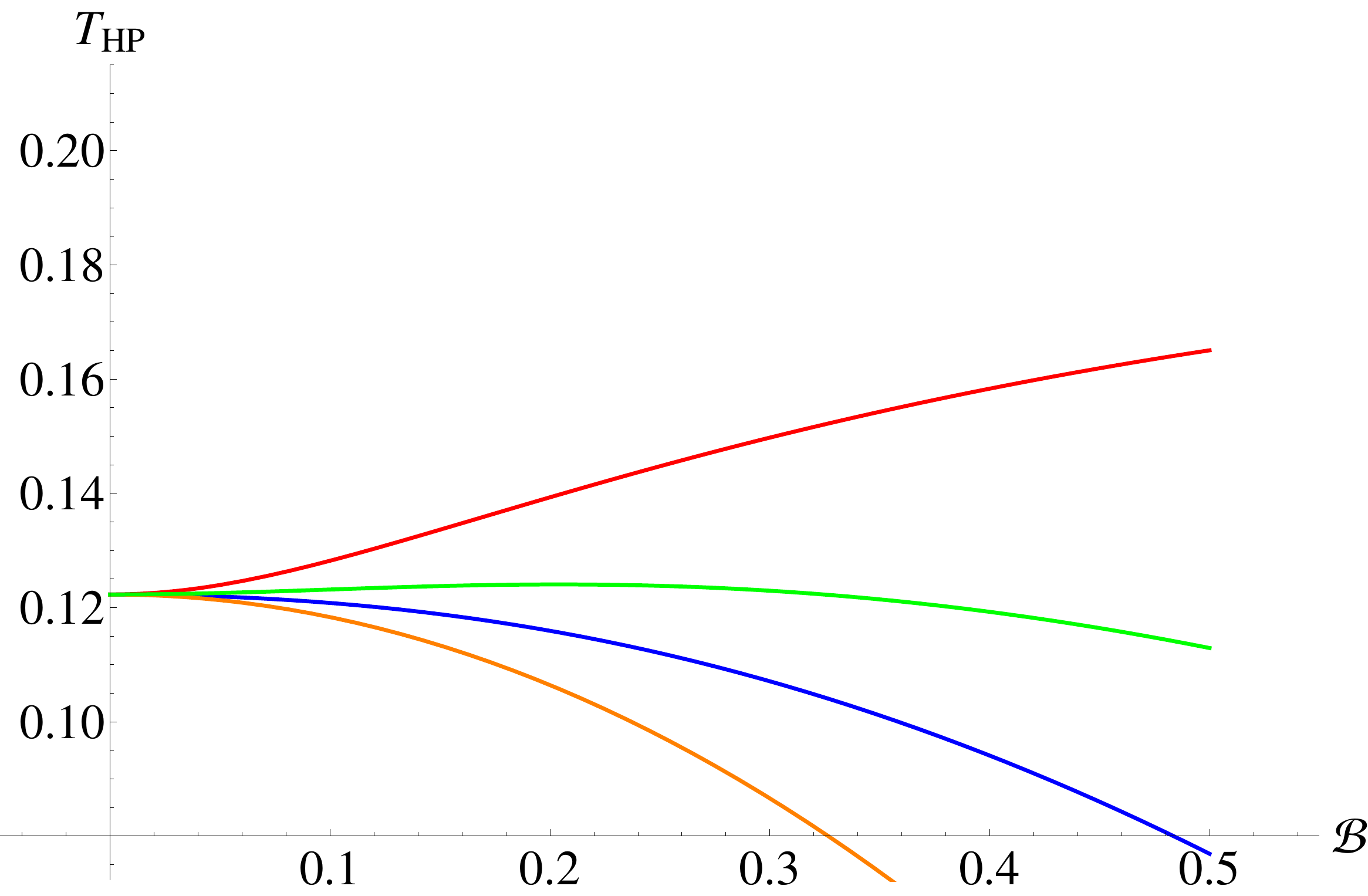}\hfill    \includegraphics[width=0.5\textwidth]{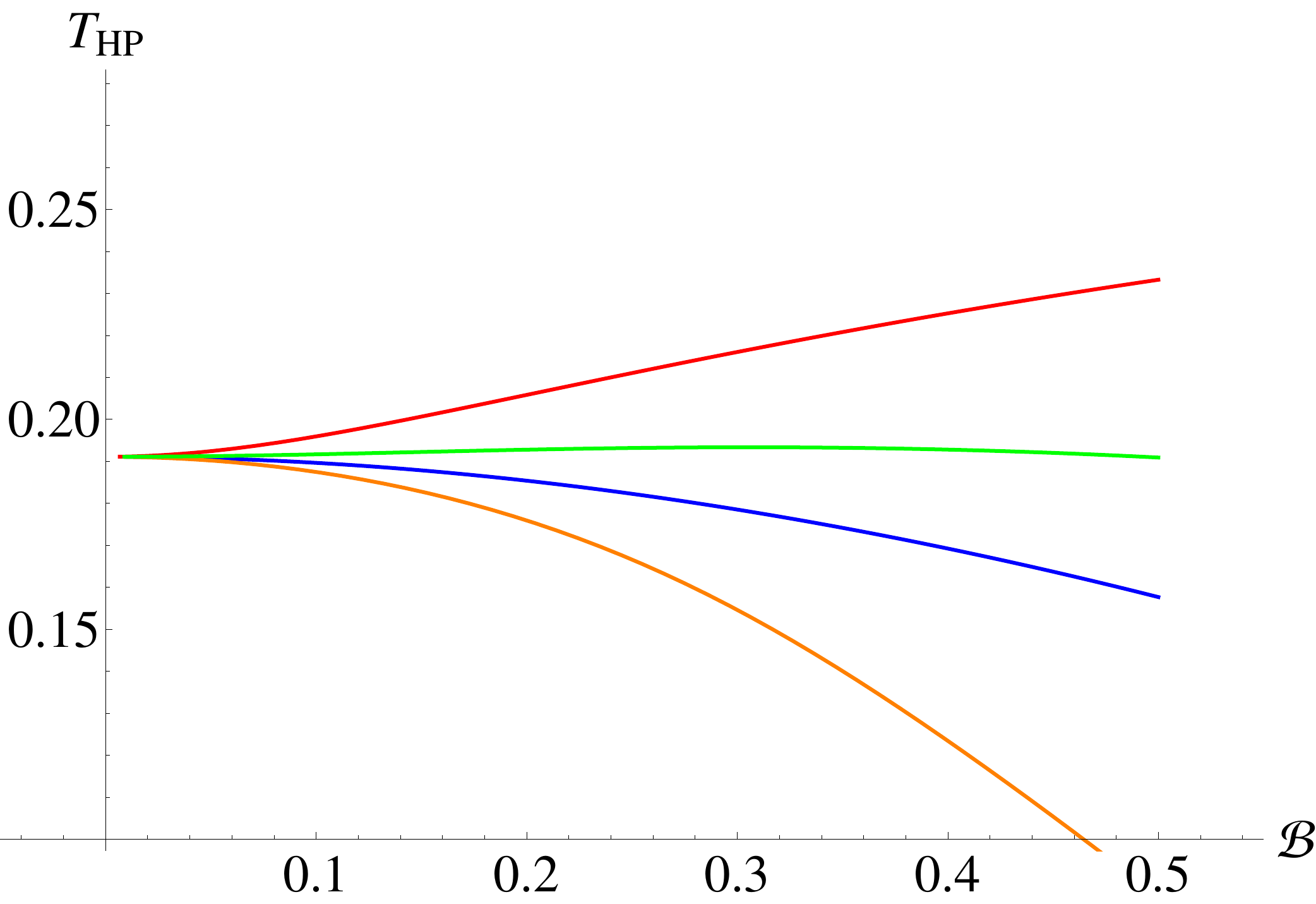}
  \caption{The Hawking-Page (deconfinement) critical temperature $T_{HP}(B)$ for $\ell_c=0.1$ GeV$^{-1}$ (red), $\ell_c=0.5$ GeV$^{-1}$ (green), $\ell_c=1$ GeV$^{-1}$ (blue), $\ell_c=2$ GeV$^{-1}$ (orange). The l.h.s.~, resp.~r.h.s.~corresponds to the hard, resp.~soft wall model.}
  \label{fig1}
\end{figure}

Let us now summarize the main results of \cite{Dudal:2015wfn}. We first analyze the Hawking-Page transition in hard and soft wall model. We computed the free energies (bulk gravity actions) $S_{conf}$ and $S_{deconf}$ using the associated metrics \eqref{thmetric}-\eqref{bhmetric} and checked at which $T$ we encounter $\Delta S=S_{deconf}-S_{conf}=0$, i.e.~at which $T$ the black hole metric becomes thermodynamically favoured. The resulting critical temperature is shown in Figure~\ref{fig1}.
We thus obtain very similar results for the hard\footnote{This case was handled before in \cite{Mamo:2015dea}, whilst omitting $\ell_c$.} and soft wall model, but we clearly notice the imprint of $\ell_c$ on the deconfinement temperature in function of $B$. Before continuing, we will thus need to discuss a way to fix $\ell_c$ in the (soft) wall model using $T=0,B\neq0$ lattice QCD data, in order to make predictions for $T\neq0,B\neq0$ data that enables comparison with output from other models or lattice simulations.

\subsection{A closer look at the chiral condensate in the soft wall model}
\begin{figure}[h]
  \centering
   \includegraphics[width=7cm,clip]{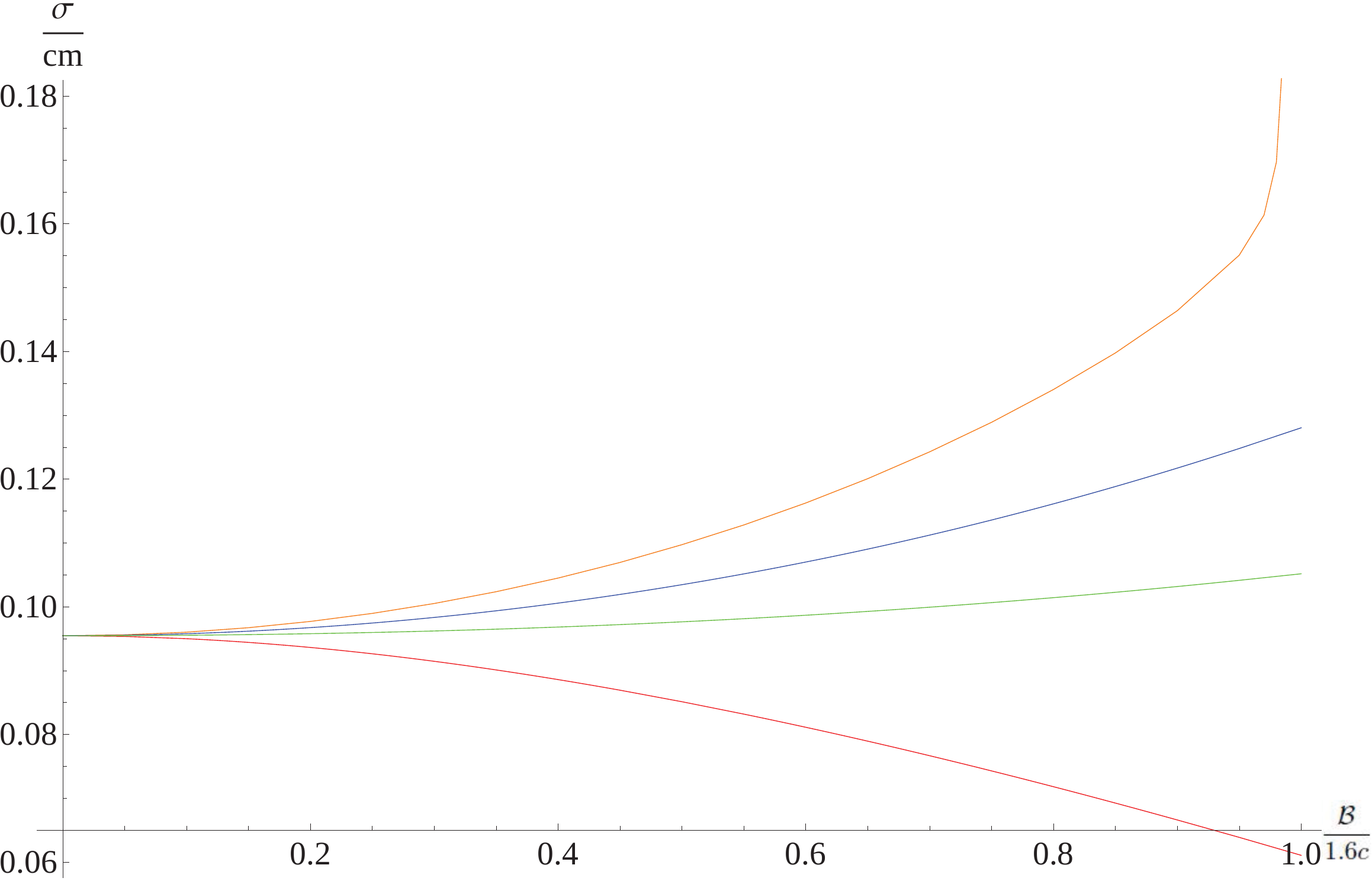}
    \caption{The ``chiral condensate'' $\frac{\sigma}{cm}$ in function of the magnetic field at $T=0$ for $\ell_c\sqrt{c}=0.1$  (red), $\ell_c\sqrt{c}=0.5$  (green), $\ell_c\sqrt{c}=1$ (blue), $\ell_c\sqrt{c}=2$ (orange).}
    \label{fig2}
\end{figure}

Roughly speaking, the AdS/CFT correspondence teaches us that
\begin{equation}
\text{Bulk field}_{r=0} = \text{source}\times r^a + \text{VEV}\times r^b+\ldots\end{equation}
where $a$, $b$ are in 1-1 correspondence with the (classical) dimension of the considered bulk field. So, in the ``chiral sector'' we indeed expect the boundary expansion written down in \eqref{rand}. Though, one rather finds in the soft wall model the boundary expansion
\begin{equation}
L^{3/2}X_0 = mr+\sigma r^3+cmr^3\ln \left(\sqrt{c}r\right)+\mathcal{O}(r^5)\,,
\end{equation}
which seems to imply that there is a logarithmic ambiguity luring in the definition of the ``chiral condensate'' $\sigma$. To overcome this, we revisited the standard quantum field theory derivation of the chiral condensate:
\begin{equation}\braket{\overline\psi\psi}=\frac{1}{Z}\frac{d Z}{d m} = \frac{\int \left[\mathcal{D} \psi \mathcal{D}\bar{\psi}\right]\left(\int d^4\mathbf{x}\bar{\psi}\psi\right) e^{-\int d^4\mathbf{x}\mathcal{L}}}{\int \left[\mathcal{D} \psi \mathcal{D}\bar{\psi}\right] e^{-\int d^4\mathbf{x}\mathcal{L}}}\,.
\end{equation}
Using holography, the path integrals in bulk and boundary are identified, so we evaluated the above to find, for $\epsilon\to0$,
\begin{equation}
\frac{16\pi^2}{N_c} \frac{m}{2}\braket{\bar{\psi}\psi} = \left(m^2 \frac{1}{\epsilon^2} + 4m^2c\log(\sqrt{c}\epsilon) + m^2c + 4m\sigma \right)\,.
\end{equation}
In principle, we need holographic renormalization to get sensible (finite) results. Though, this also can be circumvented, by subtracting the $T=0$ condensate and hence defining the chiral condensate via
\begin{equation}
\braket{\bar{\psi}\psi}_{\mathcal{B},T} =   \braket{\bar{\psi}\psi}_{\mathcal{B}=0,T=0} +\frac{N_c}{2\pi^2} \left(\sigma(\mathcal{B},T) - \sigma(\mathcal{B}=0,T=0)\right) \, .
\end{equation}
Notice that this definition is rather similar to the lattice subtracted definition of the $B$-dependent chiral condensate, see \cite{Bali:2012zg}. Doing so, we get Figure~\ref{fig2}.
Also here, there is a quite noticeable effect of $\ell_c$. Setting $\ell_c\approx 1$ GeV$^{-1}$ results in a reasonable agreement with the lattice estimation for small $\mathcal{B}$ for what concerns the slope of $\Delta\Sigma$, the subtracted condensate, see \cite{Bali:2012zg}. \\
Going back to Figure~\ref{fig1}, one finds inverse magnetic catalysis for the deconfinement transition.

 \subsection{The chiral transition}
\begin{figure}[h]
    \centering
   \includegraphics[width=7cm,clip]{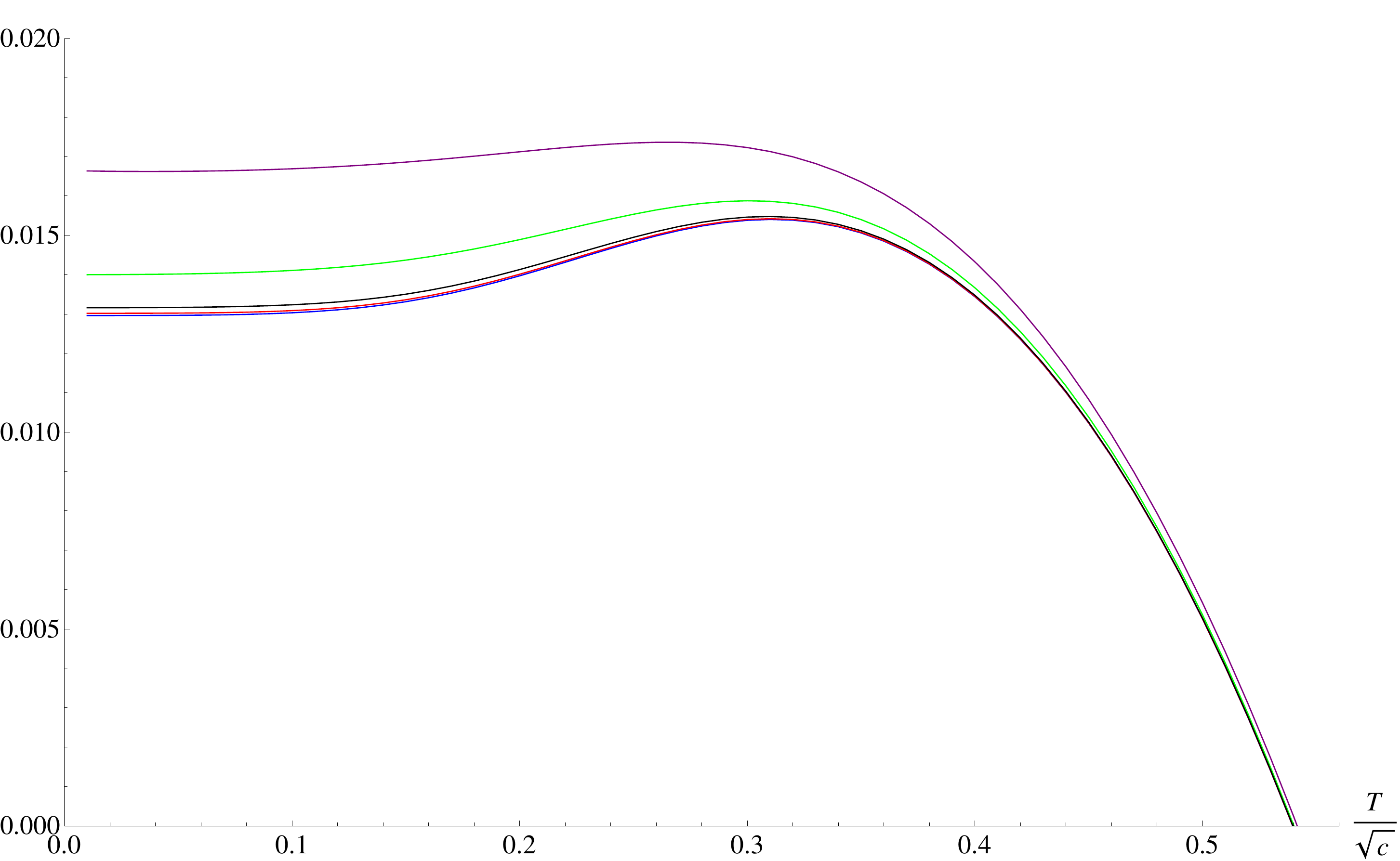}
  \caption{The ``chiral condensate'' $\frac{\sigma}{cm}$ in function in function of temperature for (from bottom to top) $\frac{\mathcal{B}}{c} = 0$ (blue), $\frac{\mathcal{B}}{c} =  0.1$ (red), $\frac{\mathcal{B}}{c} = 0.2$ (black), $\frac{\mathcal{B}}{c}= 0.5$ (green), $\frac{\mathcal{B}}{c} = 1.0$ (purple). }
  \label{fig3}
  \end{figure}

Having fixed $\ell_c$ with $T=0$, $\mathcal{B}\neq0$ lattice data, we now turn on the temperature and let the system ``decide'' what happens. This leads to Figure~\ref{fig3} for the chiral condensate.

\begin{figure}[h]
    \centering
   \includegraphics[width=7cm,clip]{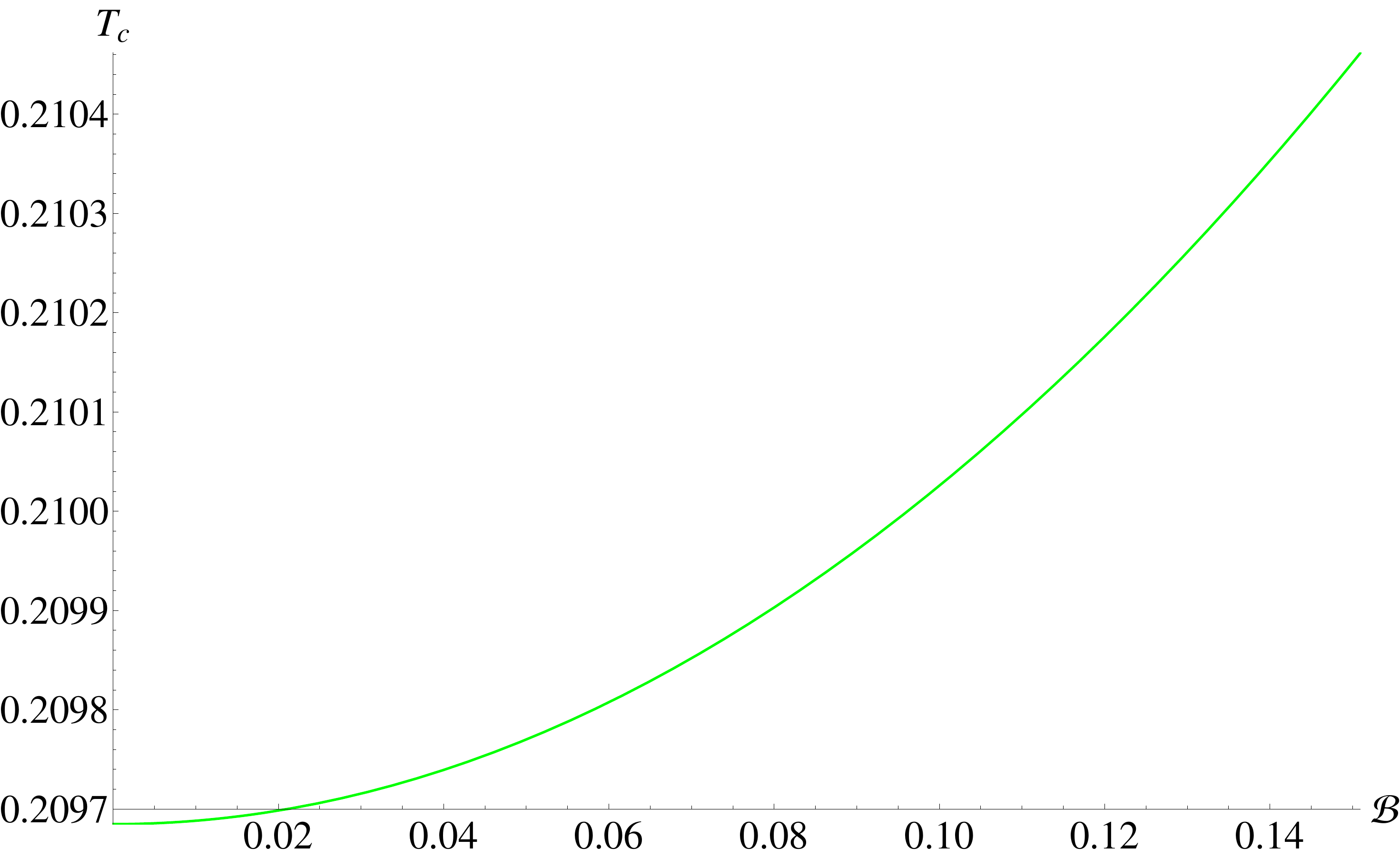}
  \caption{(in GeV) No evidence of inverse magnetic catalysis in chiral sector.}
  \label{fig4}
  \end{figure}

Finally, we consider the chiral transition temperature, as determined from
\begin{equation}
0=\braket{\bar{\psi}\psi}_{\mathcal{B},T_c} =   \underbrace{\braket{\bar{\psi}\psi}_{\mathcal{B}=0,T=0}}_{\approx -0.013~\text{GeV}^3} +\frac{N_c}{2\pi^2} \left(\sigma(\mathcal{B},T_c) - \sigma(\mathcal{B}=0,T=0)\right)\,,
\end{equation}
the result of which is shown in Figure~\ref{fig4}.

In contrast with the deconfinement transition, we can thus report no holographic evidence from our current magnetized soft wall model for inverse chiral magnetic catalysis.

\section{Discussion}
Despite the fact that during the past year or so, a few works appeared suggesting inverse magnetic catalysis can be obtained also with holographic means, our opinion is that the hunt for clear evidence of holographic inverse magnetic catalysis is still open. We already discussed the hard wall deconfinement transition in the magnetized hard wall model of \cite{Mamo:2015dea} and its inherent weaknesses.

In \cite{Li:2016gtz}, a backreacted (at leading order in $B$ and $N_f/N_c$) Sakai-Sugimoto model was constructed. Unfortunately, the fate of inverse magnetic catalysis, for the deconfinement transition in this case, is still inconclusive. One of the reported ambiguities relates to too many integration constants that corresponds to VEVs of (yet undetermined) gauge invariant QCD operators. Depending on these constants, quite different behaviour can be found.

Alternatively, in \cite{Fang:2016cnt}, the same metric as in our wall model was used, but supplemented with quark mass anomalous dimension effects in the QCD boundary expansion. As a consequence, inverse magnetic catalysis was reported. One possible caveat, to the best of our understanding, is that a wrong sign anomalous dimension seems to have been used when compared to the genuine QCD case. If so, it remains to be seen if there is a consequence at the level of the chiral transition.

In the future, we plan to further study the melting of charmonia and of (anomalous) heavy quark transport properties in a magnetized quark-gluon plasma. In this perspective, let us point out that in se, chiral (symmetry breaking) properties are less important for charmonia as charm quarks have large bare masses. As we have just reported, the  magnetized soft wall has decent deconfinement properties, so it appears to be a sensible starting point for charmonia studies, which are of more phenomenological interest than the conceptual issue of the deconfinement and chiral transition in terms of its order parameters.

In order to get a (more) consistent magnetized soft wall model with better QCD-like chiral properties, we hope to add, at some point, potentials for the dilaton $\Phi$ and scalar field $X$ (controlling chiral symmetry breaking) on top of the earlier discussed $B$-metric. The ultimate challenge would be to build, both with and without black hole, a fully self-consistent metric with a magnetic background present. A first, more moderate, step in such program could be set by following the strategy of \cite{Gherghetta:2009ac} for the $V(X)$ potential. Very recently, such a strategy was adopted in \cite{Li:2016gfn} (see also \cite{Li:2016smq}) but in our opinion, also in the latter references, (unnecessary) expansions and assumptions are made concerning the analogue of the extra length scale $\ell_d$, which is not a free scale (in contradistinction with $\ell_c$), but is related to the black hole parameters. An additional complication is that the relation between the chiral condensate and the parameter $\sigma$ becomes also clouded if higher than quadratic terms in $X$ are present in the bulk action.

\section*{Acknowledgments}
T.G.~Mertens is supported
by a postdoctoral grant of the Research Foundation-Flanders and by Princeton University.

\end{document}